\documentclass{pasj00}

\begin{document}
\SetRunningHead{T. Kouzu et al.}{Spectral Variation of the Crab Hard X-ray Emission}

\title{Spectral Variation of the Hard X-ray Emission from the Crab Nebula with the Suzaku Hard X-ray Detector}
\author{Tomomi \textsc{Kouzu},\altaffilmark{1}
Makoto S. \textsc{Tashiro},\altaffilmark{1}
Yukikatsu \textsc{Terada},\altaffilmark{1}
Shin'ya \textsc{Yamada},\altaffilmark{2}
Aya \textsc{Bamba},\altaffilmark{3}
Teruaki \textsc{Enoto},\altaffilmark{2}
Koji \textsc{Mori},\altaffilmark{4}
Yasushi \textsc{Fukazawa}\altaffilmark{5}
and
Kazuo \textsc{Makishima}\altaffilmark{6}
}

\altaffiltext{1}{
Graduate School of Science and Engineering,
Saitama University,
Shimo-Okubo 255, Sakura, Saitama 338-8570, Japan
}
\altaffiltext{2}{
High Energy Astrophysics Laboratory, Institute of Physical and Chemical Research (RIKEN),
2-1 Hirosawa, Wako, Saitama 351-0198, Japan
}
\altaffiltext{3}{
Graduate School of Science and Engineering,
Aoyama-Gakuin University,
5-10-1 Fuchinobe, Chuo, Sagamihara, Kanagawa 252-5258, Japan
}
\altaffiltext{4}{
Department of Applied Physics, Faculty of Engineering, University of Miyazaki, 1-1 Gakuen Kibana-dai Nishi, Miyazaki, 889-2192, Japan
}
\altaffiltext{5}{
Department of Physical Science, Hiroshima University, 1-3-1 Kagamiyama, Higashi-Hiroshima, Hiroshima 739-8526, Japan
}
\altaffiltext{6}{
Department of Physics, The University of Tokyo, 7-3-1 Hongo, Bunkyo-ku, Tokyo 113-0033
}

\email{(TK): kouzu@heal.phy.saitama-u.ac.jp}


\maketitle

\begin{abstract}

The Crab Nebula is one of the brightest and most stable sources in the X-ray sky.
Year-scale flux variation from the object was recently revealed in the hard X-ray band
by four satellites.
This marked the first detection of year-scale variability from pulsar wind nebulae
in the hard X-ray band.
The Crab Nebula has been observed
at least once a year for calibration purposes
with the Suzaku Hard X-ray Detector (HXD)
since its launch in 2005.
In order to investigate 
possible spectral changes as well as flux variation,
the archival data of the HXD were analyzed.
The flux variation reported by other instruments was confirmed in the 25 -- 100~keV band by the HXD in a few percent level,
but flux above 100~keV did not follow the trend in variation below 100~keV.
The hardness ratios produced utilizing the PIN and GSO sensors installed in the HXD exhibit significant scattering,
thereby indicating spectral variations in the hard X-ray. 
The spectral changes are quantified by
spectral fitting with a broken power-law model.
The difference between the
two photon indexes of the broken power-law model in harder and softer energy bands
is in the range of $<$ 2.54.
Taking into account flux variation of 6.3\% and spectral variation time-scale of a few days,
multi components of the broken
power-law-shaped synchrotron emission with different cooling times are suggested.

\end{abstract}

\section{Introduction}
\label{sec:introduction}

The Crab Nebula is one of the most famous pulsar wind nebulae (PWNe)
which is located in the center of a historical supernova
first recorded in Japan and China
in 1054 \citep{1942PASP...54...91D}.
An energetic, bright pulsar~(known as Crab pulsar=PSR~B0531+21) exists in the center of the Crab Nebula.
The Crab Nebula has been observed in the entire observable electromagnetic wavelength from radio to TeV gamma-rays.
From the non-thermal spectrum and strong polarization~(\cite{1976ApJ...208L.125W}; \cite{2008ApJ...688L..29F}),
emissions from the object are interpreted as synchrotron radiation in the X-ray band.
Emissions from the direction of the object are divided into PWN and pulsar components.
These two components can be separated not only by imaging analyses but also by phase-resolved timing analyses if we assume the
emission from the pulsar is almost the same as the pulsed emission.
Therefore, they are separately observed and discussed by many authors even in the hard X-ray band, where it is hard to get images.
The flux of the pulsar component accounts for $\sim$20\% of the total flux
with most of the rest stemming from the PWN in the hard X-ray band.

The pulsar essentially drives the electromagnetic radiation as follows:
energetic flows of electrons and positrons from the pulsar (pulsar winds) reach and interact with interstellar mediums
to induce a termination shock.
The high energy particles are believed to radiate synchrotron emission outside of the shock.
This radiation is called as PWNe.
Since the origins of the emissions of pulsars and PWNe are the rotation of magnetosphere of the
neutron star, they had been believed to be stable in principle.

Data stored in $\sim$12 years recently indicated
10 percent flux variation in 10 -- 300~keV with hard X-ray instruments \citep{2011ApJ...727L..40W}.
As reported by \citet{2011ApJ...727L..40W}, 
the pulsed flux of the Crab pulsar in
the 2 -- 100 keV band slightly decreased by 0.2\% yr$^{-1}$,
which is negligible for hard X-ray variation.
Therefore the variation seen in the Crab observations is
supposed to be due to that of the flux of PWN.
Currently, there are no reasonable models to account for the variation of the flux of PWN.
\citet{2011ApJ...727L..40W} also suggest the softening
phenomena in the light curves of $<$ 100~keV
with Fermi/GBM and INTEGRAL/ISGRI.

As indicated by \citet{2009ApJ...704...17J} and \citet{2003ApJ...598..334L},
the X-ray spectrum from the Crab Nebula is described by the broken power-law model, whose break energy is at about 100 keV.
This break can be explained by electron cooling via synchrotron radiation.
Thus the break point and the index may be changed in accordance to the flux
change. 
In fact,
\citet{2003ApJ...598..334L} reported variable broken power-law spectra with CGRO/BATSE,
while \citet{2009ApJ...704...17J} could not confirm the phenomenon.
Thus, the existence of the variation in the broken power law spectrum was unclear.
Here we analyzed the archival data of the Crab Nebula obtained by the Suzaku Hard X-ray Detector (HXD),
which offers the best sensitivity in the range of 10 -- 600~keV.

Since the promising variability could be so small compared with the systematic errors in the standard
calibration of the instruments, in principle, we have to clarify the validity of the energy response
matrix including the effective area before the spectral analyses.
This paper reports on spectral variations with close attention paid to instrumental calibration,
and is organized as follows:
Section~\ref{sec:observation} discusses the observations and the method of data reduction employed.
Section~\ref{sec:analysis} describes the results obtained.
In Section~\ref{sec:hardness},
we first examined the spectral changes without model fittings in order
to reduce possible errors from the uncertainty of the response matrices and the method of model fittings.
Then in Sections \ref{sec:indv_spec} to \ref{sec:quant},
we performed model fittings to the spectra in order to quantitate the spectral changes.
Section~\ref{sec:pulsed} describes the analysis and results regarding the pulsar component.
Finally, Section \ref{sec:discussion} discusses the origin of the observed variability.

\section{Observation and Data Reduction}
\label{sec:observation}

\subsection{Observation}

We used data observed by
the PIN (10 -- 70~keV) and GSO (Gd$_2$SiO$_5$(Ce); 40 -- 600~keV) of
the HXD (\cite{2007PASJ...59S..35T};
\cite{2007PASJ...59S..53K}) on board Suzaku~\citep{2007PASJ...59S...1M}.
The HXD has two advantages
in investigating hard X-ray spectral variability:
wide-band sensitivity in 10 -- 600~keV,
and small background rate and high reproducibility (i.e., smaller systematic errors) of the background model.
Suzaku also carries CCDs called the X-ray Imaging Spectrometer (XIS; \cite{2007PASJ...59S..23K}).
However these devices are difficult to use for this study
due to a pileup problem when observing the bright object like the Crab Nebula.
For this reason, we did not use the XIS data in this work.

There are two typical observation attitudes for Suzaku corresponding to the optical axes of the XIS and HXD
 (respectively called ``XIS nominal'' and ``HXD nominal'').
Suzaku observed the Crab Nebula with both nominal positions to check the effective areas and energy responses.
Observations for this purpose were made 17 times
from July 2005 to March 2012, and all data were public soon after the observations.
In order to avoid relative large uncertainty in calibration, 
we excluded the datasets other than nominal pointing positions; i.e., offset observations of the Crab Nebula,
which is used to define the optical axes at the initial phase of the satellite.

The effective area of the HXD is different by the nominal positions
simply due to the transparency of its fine collimator.
\citet{2007PASJ...59S..35T} and \citet{2007PASJ...59S..53K}
show that the ratio between effective areas of the HXD and XIS nominal positions is about 92\%,
and the differences in energy dependencies of the effective area between them is within 1\% in 15 -- 70~keV
and 2\% in 50 -- 600~keV.
Every response matrix is defined for a period, ``epoch'', which is divided by the change of operational parameters; combination of a bias
voltage and a set of lower discriminator (LD) level.
Table 3 of \citet{2010SPIE.7732E..81N} summarizes the history of these operations.

\subsection{Data Reduction}
\label{sec:data_reduction}

The gain changes of the PIN are within 1\% \citep{2010SPIE.7732E..81N}.
We also confirmed the PIN gain stability with  accuracy of $<$ 0.7\% until 2012 September after the study by Nishino et al (2010),
by measuring the Gd K lines produced in GSO crystals which are irradiated by X-rays from objects~\citep{kouzuphd}.
The gain of the photomultipliers for GSO changes in various timescales,
so that the GSO gain history files are constantly revised along with calibration results.
The energy scale in $<$ 100~keV was also improved by the FTOOLS \texttt{hxdpi}
\citep{2011PASJ...63S.645Y}.
All HXD data were reprocessed
with \texttt{aepipeline} version~1.0.1 in the HEADAS 6.11 software package
using CALDB 2011-09-15; therefore,
the GSO data were applied to \texttt{hxdpi} version 2010-01-12
and the gain history \texttt{gsogpt} in this work.

Background for the HXD data is generated as
a synthetic model that accounts for the time-variable particle
background (the ``non X-ray background'', NXB; \cite{2009PASJ...61S..17F}).
Specifically,
we used the ``tuned'' PIN NXB models of version 2.0 (\texttt{METHOD=LCFITDT, METHODV=2.0pin0804}).
We also used the GSO NXB models of version 2.5 (\texttt{METHOD=LCFIT, METHODV=2.5ver0912-64}) for observations before November 2011,
and those of version 2.6 (\texttt{METHOD=LCFIT, METHODV=2.6ver1110-64}) after November 2011.
The uncertainty of the NXB models of the PIN is within 3\% \citep{2009PASJ...61S..17F},
which is negligible
as the Crab source count rate is 10 times greater than that of NXB, even if at 70~keV.
The uncertainty of the GSO NXB is less than 1\% \citep{2009PASJ...61S..17F}.

In the estimation of the flux from the object, it is important to calculate the dead time (or live time) of the
observation.
The dead time of the HXD-PIN and GSO is calculated regarding the following three processes:
(1)~event processing stopped by another process running previously triggered,
(2)~an event data packet discarded in communication between the HXD analog electronics and digital electronics, and 
(3)~space packets containing events discarded due to limited bandpass between the digital electronics and data processor.
In particular, the HXD team tunes the parameters of analog electronics, 
and so the case (2) only occurs just after South Atlantic Anomaly passages.
Also, we discard events during the periods in case (3) by the FTOOL, \texttt{hxdgtigen}.
The dead time is calculated and stored with the pulse height data obtained by the FTOOL \texttt{hxddtcor}.
The dead time fraction is determined with accuracy of $<$ 0.2\%
at an observation duration of $>$ 10~ks \citep{2007PASJ...59S..53K}.

\section{Analysis of Total Emission}
\label{sec:analysis}

\subsection{Flux Variation}
\label{sec:flux}

In order to examine the time variation of total flux (pulsar + nebula) reducing effects by any systematic uncertainty of the instrumental responses, 
we evaluate the count rates of individual observations by subtracting both the non X-ray background and cosmic X-ray background from events.
We also have to note that different bias volatages for PIN were
applied during epoch 1, epoch 2 and after epoch 3, as represented in Table 3 of \citet{2010SPIE.7732E..81N}.
Effective areas for each epoch is calculated based on ground tests
(\cite{2001IEEE.TNS...48..426}; \cite{2010SPIE.7732E..81N}). 
In order to compare the measured flux in each epoch, 
here we normalized the photon flux using the calculated effective area for each epoch.
In addition to the effect of bias voltage, the effective area decreases at $<$ 25~keV because of LD of each PIN detector.
Besides,
the detector's responses above 55~keV were affected by the degraded depth of depletion layers, which is adjusted
by using the calibration data-sets on the Crab Nebula itself.
Therefore, we only used the 25 -- 55 keV band where the effective area of the response matrices changes within 1\%
for observations after PIN epoch 3.
Responses of this energy band depend on neither signal processing nor time degradation,
and are only defined simply by the physical process; cross section of the photo absorption process, or 
a stopping power of Si of the PIN diode (\cite{2010SPIE.7732E..81N}; \cite{2007PASJ...59S..53K}).
The count rates are normalized by the effective area calculated for each observation,
taking into account the difference in effective area according to different nominal positions.

\begin{figure}
  \begin{center}
    \FigureFile(100mm,60mm){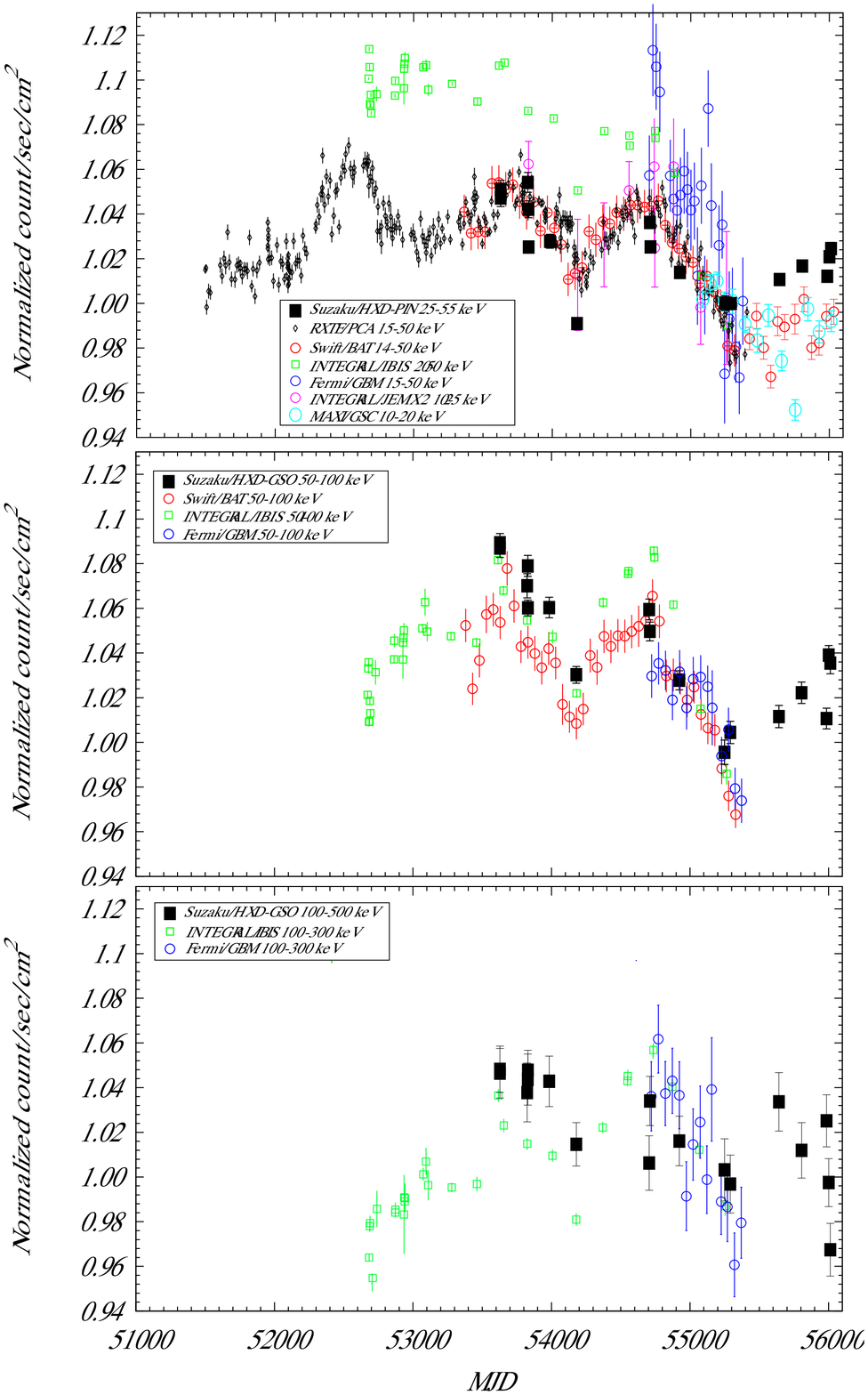}
  \end{center}
  \caption{
Long-term light curves with the Suzaku/HXD
in 25-55~keV(top), 50-100~keV (middle), and 100-500~keV (bottom)
overlaid on data of Swift, RXTE, INTEGRAL, Fermi and MAXI.
The data obtained by RXTE, Swift, INTEGRAL and Fermi before MJD 55000
are from \citet{2011ApJ...727L..40W}.
Swift after MJD 55000 and MAXI~\citep{2009PASJ...61..999M} are taken from their archival data
[(Swift) http://swift.gsfc.nasa.gov/docs/swift/results/transients/,
(MAXI) http://maxi.riken.jp/].
The error bars of the HXD include
1$\sigma$ statistical errors
and 1$\sigma$ systematic errors of NXB~\citep{2009PASJ...61S..17F}.
Each flux is normalized to mean flux in the time interval of MJD 55000--55500.
}\label{figure1}
\end{figure}

Figure~\ref{figure1} shows the variation of total count rates 
with the HXD
and other X-ray detectors
normalized with the count rates in MJD 55000 -- 55500.
The count rates in 25 -- 55~keV and that in 50 -- 100~keV
are consistent with the year-scale trends by other satellites
except for the data after MJD 55500.
On the other hand,
the trend of the GSO data in 100 -- 500~keV
deviates from trends of the other energy bands
even when the data include statistical errors ($1 \sigma$)
and systematic errors (1\% of the NXB).
In addition to the year-scale changes,
a day-scale variation of $\sim$1\% with the PIN and GSO
is also represented,
which is derived from a ratio between count rates of 2006.03.30 and 2006.04.04
whose configurations (PIN epochs and nominal positions) are the same
(see Table~\ref{table1} for the time intervals and count rates).

\subsection{Hardness Ratios}
\label{sec:hardness}

\begin{figure}
  \begin{center}
    \FigureFile(80mm,60mm){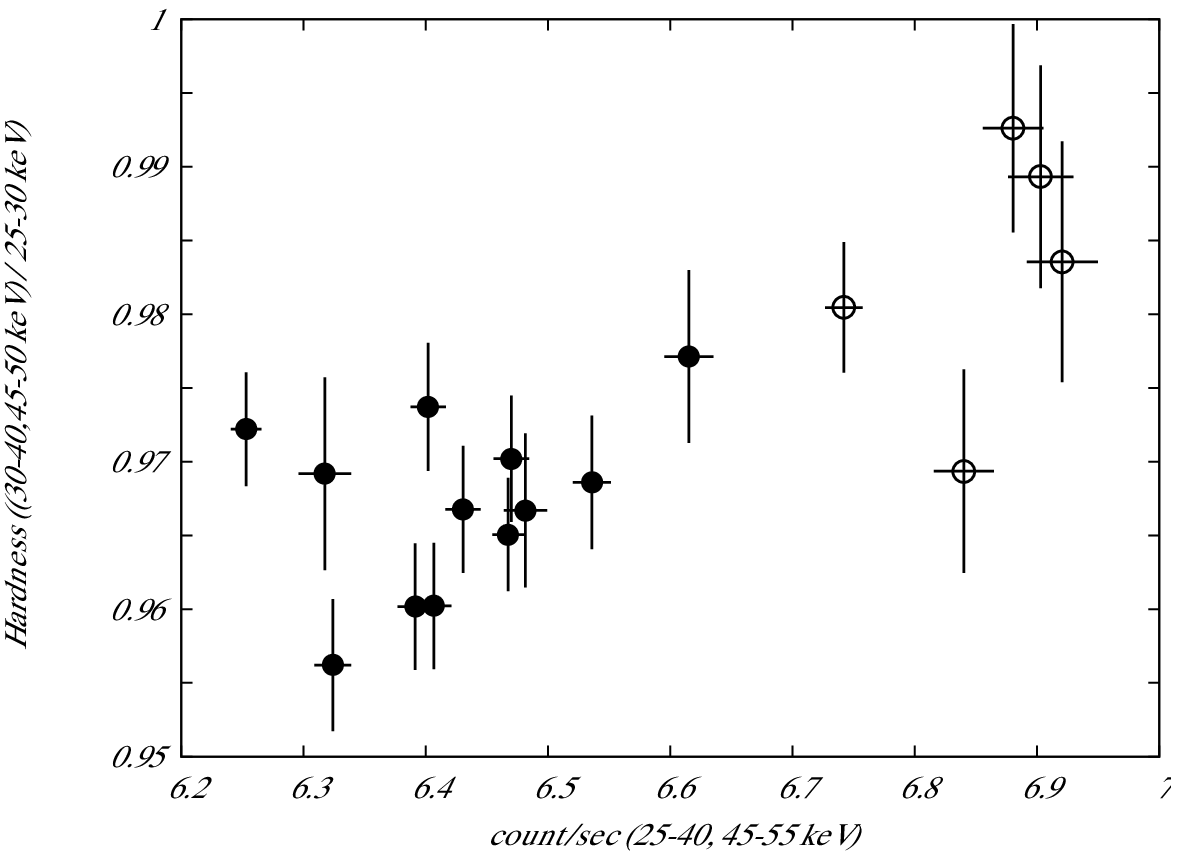}
    \FigureFile(80mm,60mm){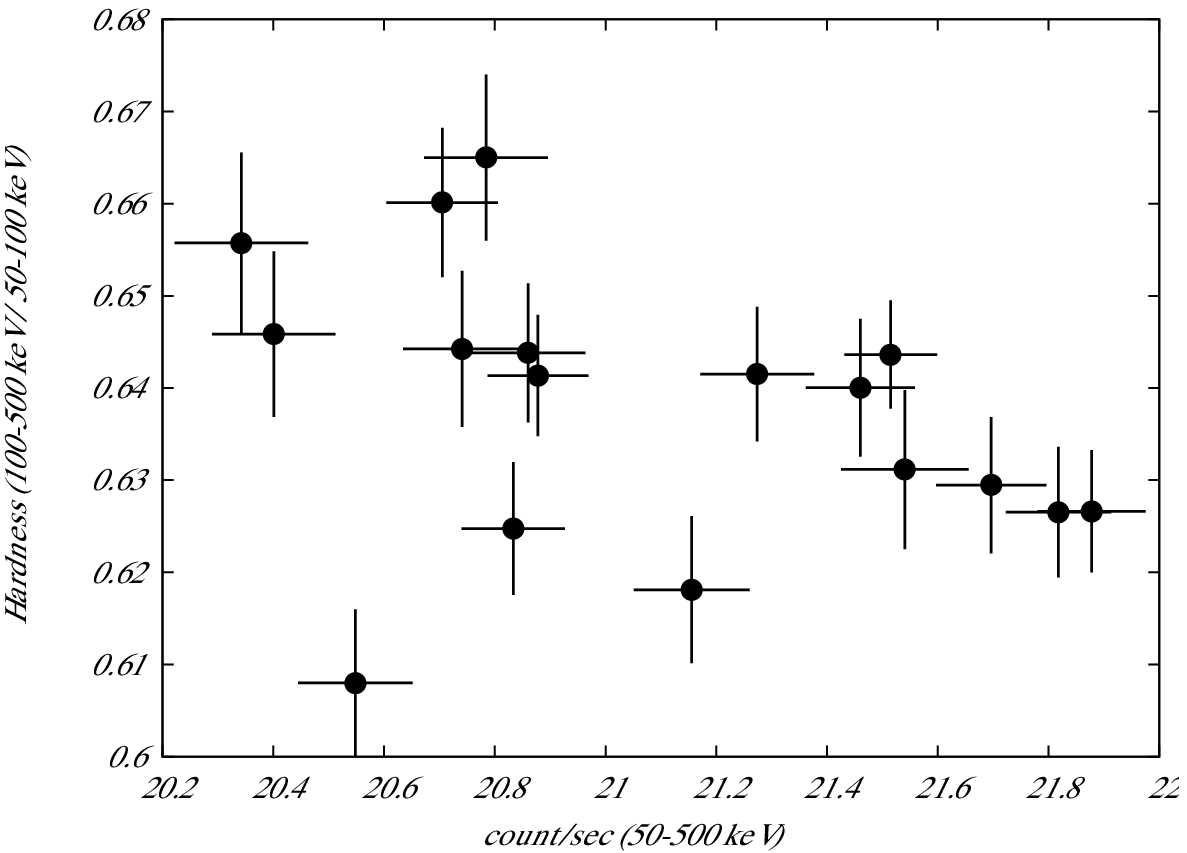}
  \end{center}
  \caption{
Count rates vs. hardness ratios of PIN 30-40, 45-50 keV/25-30 keV (left) and GSO 100-500 keV/50-100 keV (right).
The horizontal axes show count rates and the vertical axes show hardness ratios of count rates.
Data marked with open circles in the left graph are only references
because different bias voltages of the PIN were set in these observations.
The error bars represent statistical errors (1$\sigma$).
}\label{figure2}
\end{figure}

To clarify the year-scale and day-scale variations (Section \ref{sec:flux}) without model fittings,
we produced the hardness ratios of the PIN and GSO datasets.
Figure~\ref{figure2} shows the count rates vs. hardness ratios of the PIN and GSO.
The count rates (horizontal axes) are normalized by the effective area calculated for each observation,
taking into account for different nominal positions.
To avoid mixing the data of different detectors (i.e, different energy response matrices),
here we plotted hardness ratios of the PIN and GSO individually.
When observing such bright sources as the Crab Nebula, Gd K lines ($\sim$43~keV) from the GSO
had not been reproduced with sufficient accuracy in the response~\citep{2010SPIE.7732E..81N}.
We did not use the energy range of 40 -- 45 keV to avoid the Gd K
line structure.

In order to check correlations between hardness ratios and count rates,
we fitted the data with linear functions, and then got $\chi^2$/d.o.f = 16.72/10 (PIN) and 47.96/15 (GSO).
Hypotheses of linear correlations are rejected on 91.9\% (PIN) and $>$ 99.9\% (GSO) confidence levels.
In other words, both the hardness ratios and individual count rates vary significantly,
but neither shows any unique correlation.
Moreover, the hardness ratios of the soft band and hard band
behave differently,
suggesting
a variable break energy or photon indexes in spectra
which they are described by the broken power law model, as indicated by \citet{2003ApJ...598..334L}.

\subsection{Individual Spectra}
\label{sec:indv_spec}

The level of the observed variation ($\sim$ 6.3\%) of PIN 25 -- 55~keV count rates, as evaluated in section~\ref{sec:flux}, is fairly close to the sum of
official values of the systematic errors of released response matrices (gain stability), NXB,
and dead time correction ($<$ 1.2\% total at $<$ 100~keV; Section~\ref{sec:data_reduction}).
Here we summarize evidences showing that the above analysis is reliable.
As the response matrices of the HXD-PIN (25 -- 55~keV) and GSO 
are constructed only in line with Geant4 simulations with parameters obtained in ground tests
\citep{2005IEEE.TNS...52..902},
they have not been tuned or calibrated with the Crab Nebula.
Thus we reasonably consider the response matrices to be independent of the flux variation of the Crab Nebula.
As mentioned in Section 3.1,
the HXD flux trend was consistent with other satellites at 3$\pm$1\% (PIN 25 -- 55~keV), 1$\pm$1\% (GSO 50 -- 100~keV),
and 4$\pm$1\% (GSO 100 -- 500~keV), taking into account day-scale variability as the second errors (Fig.~\ref{figure1}).
These facts support
the conclusions that effective areas of the PIN 25 -- 55~keV, GSO 50 -- 100~keV and 100 -- 500~keV are essentially stable within 4\%, 2\% and 5\% at most
respectively,
and that the relative flux variation is real.

Before evaluating individual spectra,
we first performed a model fitting of the averaged spectrum of all HXD observations in 25 -- 40, 45 -- 55~keV
as a template to evaluate spectral variations,
using the averaged response matrix weighted by individual exposures.
We used a single power-law model 
in order to see possible spectral break in comparison
with the simple model.
The values of the photon index and flux are 2.141 $\pm$ 0.006 and 9.65$^{+0.1}_{-0.2} \times10^{-9}$ erg cm$^{-2}$ s$^{-1}$
(25 -- 55 keV)
at a 90\% confidence level with
the $\chi^2$/d.o.f. = 68.54/63.

\begin{figure}
  \begin{center}
    \FigureFile(140mm,100mm){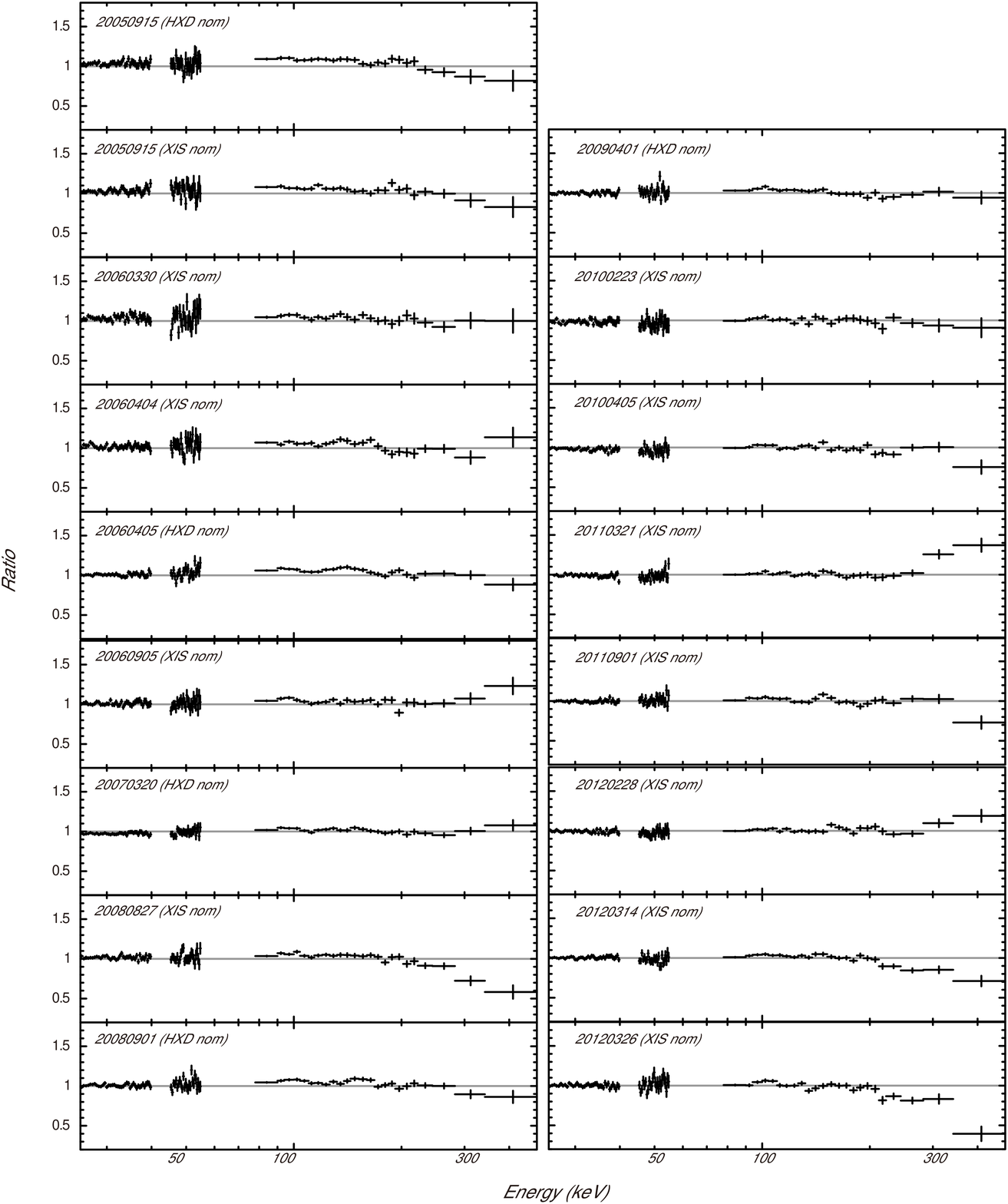}
  \end{center}
  \caption{
Ratios of background-subtracted spectra of individual observations to the best-fit single power-law model of the averaged PIN spectrum
(25-40, 45-55~keV) of all observations.
The error bars represent statistical errors (1$\sigma$).
}\label{figure3}
\end{figure}

Second, we compared individual spectra with this model.
We used the appropriate arf files (\texttt{ae\_hxd\_gsohxnom\_crab\_20100526.arf} or
\texttt{ae\_hxd\_gsoxinom\_crab\_20100526.arf})
for all the GSO spectra,
in addition to the response matrices (\texttt{ae\_hxd\_gsohxnom\_20100524.rsp} or
\texttt{ae\_hxd\_gsoxinom\_20100524.rsp})
for the aiming positions (HXD nominal and XIS nominal).
The GSO arfs are empirical functions
to correct the normalization of the GSO for
matching that of the PIN,
implemented as spline functions converging to a constant above 100 keV
\citep{2011PASJ...63S.645Y},
and the reference
therein\footnote{http://heasarc.gsfc.nasa.gov/docs/suzaku/analysis/gso\_newarf.html}.
Figure~\ref{figure3} shows the ratios of individual spectra to the averaged best-fit power-law model.
This figure clearly demonstrates the variations of the spectral break at $\sim$100 -- 200~keV.
For instance, the spectral slope was steep in the spectra of 2005.09.15, 2008.08.27, and 2012.03.14/26,
while flat in the spectra of 2007.03.20 and 2009 -- 2010.

\subsection{Quantitative Analysis of the Averaged Spectrum}
\label{sec:avg_spec}

The hard X-ray spectrum (around $\sim$100 keV) of the Crab Nebula
is well represented by a broken power-law model (e.g., \cite{2011PASJ...63S.645Y}),
although \citet{2009ApJ...704...17J} reported that
a log parabolic law (\cite{2000A&A...361..695M};
\cite{2004A&A...413..489M}) is better fitted with the spectrum rather than a broken power law.
We thus tested several models for the hard X-ray spectrum
using the HXD data.
We averaged all spectra of the observations listed in Table~\ref{table1}.
We performed fittings with a single power law 
\begin{equation}
A(E) = KE^{\Gamma},
\end{equation}
an exponential cutoff power law, 
a log parabolic law\footnote{
We installed and used the \texttt{logpar} model
as downloaded from http://heasarc.gsfc.nasa.gov/xanadu/xspec/models/logpar.html},
the Band function \citep{1993ApJ...413..281B}, 
and a broken power-law model 
\begin{eqnarray}
A(E)  &= KE^{\Gamma_1} (E < E_{break}),\\
         &= KE_{break}^{\Gamma_1 - \Gamma_2} \left( \frac{E}{1{\rm ~keV}} \right) ^{\Gamma_2} \, (E \geq E_{break})
\end{eqnarray}
to the averaged spectrum.
We adopted the average responses calculated according to the exposure time of each observation mode (attitude and PIN epoch).
Here we note that the spatial extension of the Crab Nebula ($\sim$ 1 arcmin in diameter in hard
X-ray band; \cite{1987ApJ...319..416P}) can be negligible due to the HXD angular response
whose field of view is $34'$ $\times$ $34'$ ($<$100~keV) and 4.5\degree $\times$ 4.5\degree ($>$100~keV) (FWHM)
exhibit flat top effective areas within $\sim$ 2 arcmin in diameter.

Figure~\ref{figure4} shows the averaged spectrum and ratio of the spectrum to the best-fit models.
Table~\ref{table2} lists the best-fit parameters.
The single power-law model shows large residuals at $>$ 100~keV
to imply that a cutoff or break is necessary in this energy band.
Although the model of the second smallest $\chi^2/$d.o.f is the cutoff power-law model,
the best-fit cutoff energy is much larger than the energy band of the HXD data.
The broken power-law model,
which is an empirical model for the Crab hard X-ray spectrum,
succeeded to represent the data with the smallest $\chi^2$ among the five models above 100 keV range.
We still see a residual structure in 25 -- 55 keV band.
This may suggest spectral variation in the observations.

\subsection{Quantitative Analysis of Individual Spectra}
\label{sec:quant}

In order to evaluate the spectral variation quantitatively,
we performed model fittings for individual spectra with the broken power-law models,
which provides the smallest $\chi^2$ for the averaged spectrum (Section~\ref{sec:avg_spec}).
As shown in Figure~\ref{figure5} and Table~\ref{table3},
all the spectra are reproduced by the broken power law model
and we also note that the residual structure seen in the averaged
spectra is disappeared in these time resolved spectra.
Figure~\ref{figure6} shows the distribution of derived $\Gamma_1$ and $\Gamma_2$,
which are photon indexes below and above the break energy, respectively.
Figure~\ref{figure7} represents the derived break energy ($E_{break}$) from each observation.
In Figure~\ref{figure6} and Figure~\ref{figure7},
we evaluate the errors of the best fit values
including systematic errors due to the reproducibility of
the GSO NXB, by varying the derived NXB models by $\pm 1$\%.
Hypotheses that $\Gamma_1$, $E_{break}$ and $\Gamma_2$ were constant
are safely rejected at $>$99.9\%, 99.7\% and 91.7\% confidence levels, respectively, taking into account both statistic and systematic errors.
The very conservative upper limit of $\Gamma_1 - \Gamma_2$ is 2.54
considering the upper limit of $\Gamma_1$ and the lower limit
of $\Gamma_2$ at the 90\% confidence range.
The upper limit of $\Gamma_1 - \Gamma_2$ will be reduced $<$ 1.29 if we exclude the data point at 2006.03.30,
whose $\Gamma_1 - \Gamma_2$ is exceptionally large, 2.54, mainly because the value of $\Gamma_1$ allows $\Gamma_1=0$.
We fitted $\Gamma_1 - \Gamma_2$ with a constant as a function of MJD,
and found that the best-fit constant value is 0.16 $\pm$ 0.02 at a 1$\sigma$ error with $\chi^2$/d.o.f. = 23.74/16.
In Fig.~\ref{figure6} and Table~\ref{table3}, 
the errors increase by time because the level of NXB is increased due to accumulated events from radio active nuclei by cosmic-rays in GSO crystals.  

We also examined an exponential cutoff power-law model for the spectrum.
Figure~\ref{figure8} shows the best-fit photon indexes and cutoff energies ($E_{cut}$).
We cannot obtain conclusive results with this model,
as all best-fit values for $E_{cut}$ fairly exceed the energy range of the HXD data (500~keV).

\section{Analysis of the Pulsed Component}
\label{sec:pulsed}

\subsection{Pulse Profiles}

To understand how the pulsed component of the pulsar affects
the variation of total emission from the Crab Nebula,
we investigated the fluctuations
of pulse profiles and pulsed flux of the Crab pulsar.

We adopted the pulse period ($P$) and its time-derivative ($\dot{P}$)
from the Jodrell Bank Crab Pulsar Monthly Ephemeris\footnote{
http://www.jb.man.ac.uk/\textasciitilde pulsar/crab.html}.
Epoch (phase=0) was defined as the arrival time of the first peak in the radio band. 
We used \texttt{aebarycen} \citep{2008PASJ...60S..25T}
to correct photon arrival times at UTC in orbit to those at the solar system barycenter,
assuming that the Crab pulsar
is located in the position of R.A. = \timeform{5h34m31.97232s},
Decl. = \timeform{+22D00'52.0690''} (J2000),
on which is the same coordinate used by
the Jodrell Bank Crab Pulsar Monthly Ephemeris.

Figure \ref{figure9} shows
the pulse profiles observed with the HXD at the HXD nominal pointing position shown in count~s$^{-1}$.
Here, we defined On and Off phases at phase 0.9 -- 1.5 and 0.55 -- 0.85, respectively.
The averaged counts at the Off phase are subtracted in the plot.
The relative intensity of the first pulse peak
with respect to the second one increases
with the energy.
The X-ray pulse peak leads to
the first radio pulse peak (phase $=$ 1 in Fig.\ref{figure9})
by $\sim300$~$\mu$s,
which is consistent with results previously reported by other satellites
~(\cite{2004ApJ...605L.129R}; \cite{2003A&A...411L..31K})
and the initial observations with the HXD~\citep{2008PASJ...60S..25T}.
The pulse profiles were consistent with each other within $2.7 \sigma$ of statistical errors of each phase bin
through all observations in 25 -- 300~keV.

\subsection{Pulsed Flux}

We searched for variations of the pulsed flux
using all data at both nominal pointing positions.
Count rates of the pulsed components, ``On$-$Off'' are plotted in Figure~\ref{figure10}.
Numerically, the best fit values in fitting of data in Figure~\ref{figure10} with a constant model are
$1.68 \pm 0.01$ count s$^{-1}$ and $\chi^2/$d.o.f. = 23.03/16  at PIN 25 -- 40, 45 -- 55~keV,
$6.42 \pm 0.04$  count s$^{-1}$ and  $\chi^2/$d.o.f. = 38.96/16 at GSO 50 -- 300~keV
with the errors at 90\% confidence level.
Hypotheses for the constants are rejected at 88\% and 99\% confidence levels, respectively.
The pulsed component contributes only about 20\% of the total flux, and thus, the possible  $\sim 1$\% of variations are 
hard to account for the 
total variation of the hard X-rays shown in Fig.~\ref{figure1} with the fluctuation of the pulsed component.

\section{Discussion}
\label{sec:discussion}

In Section \ref{sec:flux},
we demonstrated the long-term photon flux
variation measured by the HXD, which is consistent with those
measured by other X-ray satellites within 3 -- 4\%
(see also Section \ref{sec:indv_spec}).
The relative peak-to-peak amplitude of the flux variation was 6.3\% in 25 -- 55~keV.
As indicated in the hardness ratio (Section~\ref{sec:hardness}),
we evaluated the spectral variation quantitatively (Section~\ref{sec:quant})
to find that the difference between the photon indexes below and above the break energy varies.
Because the flux of the pulsed component contribute only $\sim$ 20\% of the total flux (Section~\ref{sec:pulsed}), the variation seen in the total flux may be caused by
the nebular component (the Crab Nebula).
The following sections
discuss
the origin of the variation of X-ray fluxes and energy spectra.

\subsection{Comparison with Previous Works}

Our results obtained by the Suzaku HXD confirmed
year-scale flux variation reported by \citet{2011ApJ...727L..40W}
using RXTE, INTEGRAL, Swift and Fermi.
They also claimed softening in MJD
54690 -- 55390 in flux decreasing at $<$100~keV.
Our results agree with their arguments regarding the general trend during this time interval,
although the HXD results do not show simple softening in detail (Fig.~\ref{figure2}).
In the range of 100 -- 300~keV,
they reported less decrease than in the softer band in MJD 54690 -- 55390
with INTEGRAL/ISGRI.
We observed similar phenomena in the HXD data,
which showed less flux variation in the 100 -- 500~keV than those of the 25 -- 55 and 50 -- 100~keV bands.
However, the PIN count rate seems to exhibit slightly larger values than those of other instruments after MJD 55500 by about 3 \% (Fig.~\ref{figure1}). This may be due to secular degradation of charge transfer efficiency in PIN detector. In order to examine the possibility, we checked the PIN gains using Gd K line from GSO crystal and found the PIN gains in this duration might be reduced
by $-0.5 \pm 0.7$ \%. Assuming the Crab spectrum this corresponds to
the tolerance of $-1.7^{+3.4}_{-2.5}$ \% in the photon flux, which
can explain the apparent discrepancy.
But we stress that 
a hypothesis that $\Gamma_1$ values were constant is rejected at a $>$ 99\% confidence level,
even if we omit the data after MJD 55500
during which the PIN gain might have been degraded.

\subsection{Origin of the Variable Component}

The X-ray radiation of the Crab Nebula is generally regarded as synchrotron radiation
because of the non-thermal spectrum and the strong polarization.
From our analysis,
the averaged hard X-ray spectrum is represented by the broken power-law model
with $E_{break} \sim 100$~keV (Section~\ref{sec:avg_spec}).
In the PWN,
electrons are re-accelerated and continuously injected from the termination shock,
and then cooled down via synchrotron radiation.
Assumed synchrotron radiation, the higher energy electrons
are cooled in the shorter cooling time ($\tau_c$).
According to the nomenclature of Sari et al. (1998),
when $\tau_ c$ of the minimum Lorentz factor electrons is
shorter than a duration ($t_0$) for each spectrum radiation,
which presumably corresponds to the electron passage time
through the emission region, the regime is called ``fast cooling''
regime. If it is the case of ``fast cooling'', the observed
$E_{break}$ corresponds to either of $\nu_ a$, $\nu_ c$ or $\nu_ m$,
and the observed $\Gamma_1$ is either $1$, $-2/3$, or $-3/2$. However,
none of those values are accepted by any of the measured
$\Gamma_1$ except for 2006.03.30 (Table 3), meaning the observed spectra reject the
``fast cooling'' regime. We note that $\Gamma_1$ of 2006.03.30 has a large error as $E_{break}$ cannot be well derived.
On the other hand, in the case of the
``slow cooling'' regime ($t_0 < \tau_ c$), $\Gamma_1$ should be
either $1$, $-2/3$, or $-(p+1)/2$, and $E_{break} = \nu_a $, $\nu_ m$,
or $\nu_c$, respectively. The observed values of $\Gamma_1$
accept only the case of $E_{break} = \nu_ c$ in the ``slow cooling'' regime. 

Although
$\Delta \Gamma = \Gamma_1 - \Gamma_2$ should be 0.5 in the slow cooling regime since $\Gamma_2$ should be $-(p+2)/2$,
14 out of 17 data of ours would not accept the expected value assuming a simple one-zone model ($\Delta \Gamma$ = 0.5).
However here we have to notice that significant spatial and temporal varations are reported in the soft X-ray band
(\cite{1999ApJ...510..305G}; \cite{2002ApJ...577..49L}; \cite{2004IAUS..218..181M}),
which naturally implies multicomponent variable spectra also in the hard X-ray band.
Moreover the filamentary structures seen in the Chandra images (\cite{2002ApJ...577..49L}; \cite{2004ApJ...609..186M})
suggest that magnetic fields are different by locations in the Crab Nebula.
On the other hand, the photon indexes of X-ray spectra are about 2 anywhere in the Crab Nebula~\citep{2004ApJ...609..186M},
energy spectra of re-accelerated electrons are thought to have almost constant indexes.
Considering these facts, the multicomponent spectra are observed as an ensemble of power-law components and broken power-law components (with $\Delta \Gamma$ = 0.5)
in the limited observed energy band.
Consequently, we observe variable broken power-law spectra with 0 $<$ $\Delta \Gamma$ $<$ 0.5 as superpositions of these components in the energy band.
In fact, all of our data accept the expected range of 0 $<$ $\Delta \Gamma$ $<$ 0.5.

Magnetic field strength $B$ related to synchrotron radiation is given as:
\begin{equation}
\label{eq:synch_b}
\frac{B}{100~\mu {\rm G}}  \sim \left( \frac{\tau_{c}}{1~{\rm year}} \right) ^{-2/3} \left( \frac{E_{ph}}{100~{\rm keV}} \right) ^{-1/3}
\end{equation}
where $E_{ph}$ is the photon energy.
The synchrotron bolometric intensity
($P \propto B^2 n(\gamma) \gamma^2$)
is assumed to be proportional to the hard X-ray flux,
where $n(\gamma)$ is the electron number density as a function of the Lorentz factor of electron $\gamma$.
In the slow cooling regime,
$\tau_{c}$ should be comparable with the timescale of passages of electrons
that are pertinent to $\nu_c$.
From our results, the timescale of the flux variation is expected to be a few days (Section~\ref{sec:flux}).
By adopting a timescale of five days, for example,
we can estimate the value of $B$ to be $\sim$1.7 mG from Equation~\ref{eq:synch_b}.
Because the cooling timescale can be regarded as the passage time scale,
the typical scale of the variable emission region is estimated to be around 0.35 -- 1.25~light-days ($\sim$(3.2 -- 8.1)$\times 10^{-4}$~pc),
under the assumption of the speed of plasma in the PWN measured with Chandra (0.07 -- 0.25$c$; \cite{2004ApJ...609..186M}).
Although the scale of the variable region is $\sim$ 0.1\% of the whole X-ray emission region,
the mismatch between the occupancy (about 0.1\% of the whole X-ray emitting region) and the factor of the flux variations (8\% of total flux) can be solved
when the variable region has $\sim$ 10 times higher magnetic field strength than those in others, because $P \propto B^2$ as mentioned before.
Of course, because $P$ has a dependency with $n(\gamma)$ or $\gamma$, we may assume that a multiple component has a different density or Lorentz factor, but
we do not discuss these possibilities here due to limitation of information which we have currently.
We must wait for hard X-ray images to conclude the entire morphology of the variable emission region and the magnetic fields.

\bigskip


We would like to acknowledge Takayuki Yuasa, 
Shinpei Shibata and Yasuyuki Tanaka for their supports in analysis and discussion.
We would like to thank the anonymous referee for the valuable comments and suggestions,
which significantly improved our paper.
We wish to thank all members of the Suzaku operation team and
the Suzaku HXD team.
This research has made use of the MAXI data provided by RIKEN, JAXA and the MAXI team.
T.K is supported by the Grant-in-Aid for Research Fellowship for Young Scientists (DC2)
of the Japan Society for the Promotion of Science (JSPS)
No.~239311, and in part by the Grants-in-Aid for Scientific Research (B) from the Ministry
of Education, Culture, Sports, Science and Technology (MEXT) (No. 23340055, Y.T; No. 22340039, M.S.T).


\onecolumn



\begin{table}
  \caption{Suzaku Observations of the Crab Nebula}\label{table1}
  \begin{center}
    \begin{tabular}{lllllll}
      \hline
      Obs. ID & Date & MJD & Exposure\footnotemark[$*$] & Nominal position & PIN epoch & Count rate\footnotemark[$\dagger$] \\ \hline
      100023010 & 2005/09/15 & 53628 &  9.9 ksec & HXD & 1 & $7.141 \pm 0.028 \pm 0.003$\footnotemark[$\S$] \\
      100023020 & 2005/09/15 & 53628 & 12.4 ksec & XIS & 1 & $7.115 \pm 0.026 \pm 0.004$\footnotemark[$\S$] \\
      101004010 & 2006/03/30 & 53824 &  9.1 ksec & XIS & 1 & $7.162 \pm 0.030 \pm 0.004$\footnotemark[$\S$] \\
      101004020 & 2006/04/04  & 53829 & 12.5 ksec & XIS & 1 & $7.080 \pm 0.026 \pm 0.003$\footnotemark[$\S$] \\
      101003010 & 2006/04/05  & 52830 & 29.1 ksec & HXD & 1 & $6.965 \pm 0.016 \pm 0.003$\footnotemark[$\S$] \\
      101010010 & 2006/09/05  & 53983 & 18.3 ksec & XIS & 2 & $6.983 \pm 0.021 \pm 0.004$\footnotemark[$\S$] \\
      102019010 & 2007/03/20 & 54179 & 40.5 ksec & HXD & 3 & $6.733 \pm 0.013 \pm 0.003$ \\
      103007010 & 2008/08/27 & 54705 & 30.3 ksec & XIS & 4 & $7.040 \pm 0.016 \pm 0.003$  \\
      103008010 & 2008/09/01  & 54710 & 31.6 ksec & HXD & 4 & $6.966 \pm 0.015 \pm 0.003$ \\
      104001010 & 2009/04/02  & 54923 & 31.3 ksec & HXD & 5 & $6.888 \pm 0.015 \pm 0.003$ \\
      104001070 & 2010/02/23 & 55250 & 15.1 ksec & XIS & 8 & $6.795 \pm 0.022 \pm 0.004$ \\
      105002010 & 2010/04/05  & 55291 & 31.3 ksec & XIS & 9 & $6.794 \pm 0.016 \pm 0.004$ \\
      105029010 & 2011/03/21  & 55641 & 34.2 ksec & XIS & 11\footnotemark[$\ddagger$] & $6.867 \pm 0.015 \pm 0.004$ \\
      106012010 & 2011/09/01  & 55805 & 33.9 ksec & XIS & 11 & $6.908 \pm 0.015 \pm 0.003$\\
      106013010 & 2012/02/28  & 55986 & 34.0 ksec & XIS & 11\footnotemark[$\ddagger$] & $6.876 \pm 0.015 \pm 0.003$ \\
      106014010 & 2012/03/14  & 56001 & 42.2 ksec & XIS & 11\footnotemark[$\ddagger$] & $6.936 \pm 0.013 \pm 0.003$ \\
      106015010 & 2012/03/26  & 56012 & 22.9 ksec & XIS & 11\footnotemark[$\ddagger$] & $6.961 \pm 0.018 \pm 0.003$\\
      \hline
      Total &&& 437.7 ksec && \\ \hline
\multicolumn{6}{@{}l@{}}{\hbox to 0pt{\parbox{160mm}{\footnotesize
 \par\noindent
 \footnotemark[$*$] Dead time corrected exposure time.\\
 \footnotemark[$\dagger$] Count rate of the PIN in units of count s$^{-1}$ @ 25--55~keV. Errors are statistical and systematic errors of the NXB (3\%)
on 1$\sigma$ levels. Each count rate of the XIS nominal is adjusted to the HXD nominal.\\
 \footnotemark[$\ddagger$] Unusual values of the PIN epoch are selected given the special PIN LD setting during observations.\\
 \footnotemark[$\S$] Count rates of Epoch 1 and 2 are normalized to Epoch 3.
 }\hss}}
   \end{tabular}
  \end{center}
\end{table}


\begin{figure}
  \begin{center}
    \FigureFile(140mm,100mm){FIG4.eps}
  \end{center}
  \caption{
The top panel shows the averaged spectrum through 2005--2012 with the Suzaku HXD.
The solid line represents the best-fit single power-law model and
crosses are data with 1$\sigma$ statistical errors.
Below five panels show the ratios of the data to best-fit models of
a single power law (power),
an exponential-cutoff power law (cutoffpl),
a log parabolic law (logpar),
Band function (grbm),
and a broken power law (bknpower).
}\label{figure4}
\end{figure}

\begin{figure}
  \begin{center}
    \FigureFile(140mm,100mm){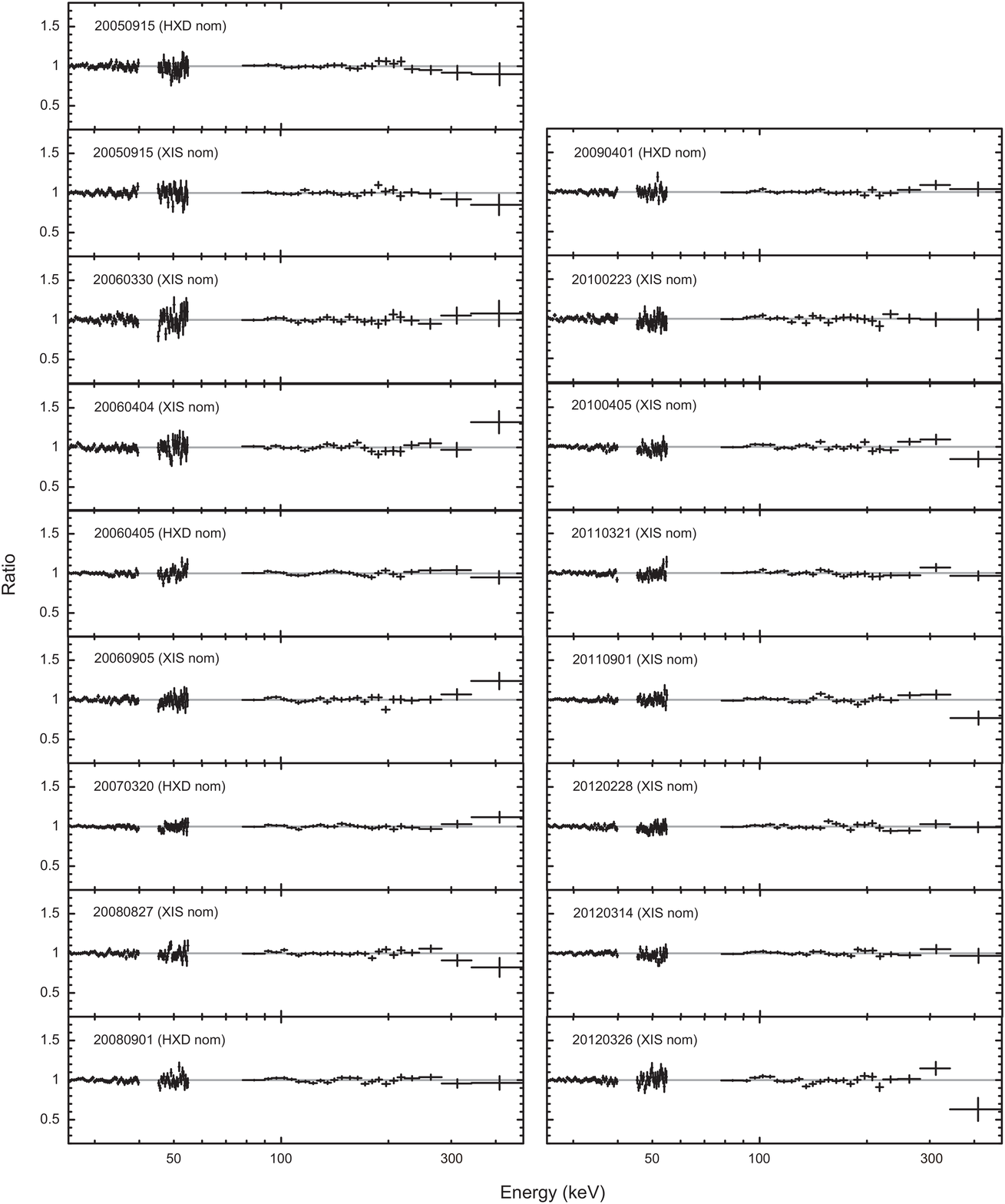}
  \end{center}
  \caption{
Ratios of data to best-fit broken power-law models of individual spectra
}\label{figure5}
\end{figure}

\begin{table}
  \caption{Best-fit models of the averaged spectrum}\label{table2}
  \begin{center}
    \begin{tabular}{lll}
      \hline
      \hline
      Model name & Parameter name & Value (error) \\
      \hline
      Single power law & $\Gamma$ & $-2.125\pm0.001$ \\
                                  &  Flux$_{s}$\footnotemark[$*$] & $0.970 \pm 0.009$ \\
                                  &  Flux$_{h}$\footnotemark[$\dagger$] & $7.865_{-0.007}^{+0.008}$ \\
                                  &  $\chi ^2 /$ d.o.f. & 797.16 / 87 \\
      \hline
      Cutoff power law & $\Gamma$ & $-2.075\pm0.004$ \\
                                  &  $E_{cut}$~(keV)  & $1396_{-113}^{+134}$ \\
                                  &  Flux$_{s}$\footnotemark[$*$] & $0.9728_{-0.0010}^{+0.0008}$ \\
                                  &  Flux$_{h}$\footnotemark[$\dagger$] & $7.951_{-0.014}^{+0.008}$ \\
                                  &  $\chi ^2 /$ d.o.f. & 431.04 / 86 \\
      \hline
      Log parabolic law & $\alpha$ & $-2.150\pm0.003$ \\
                                  &  $\beta$  & $-0.059 \pm 0.006$ \\
                                  &  $_{pivot}E$~(keV)  & 20 (fix) \\
                                  &  Flux$_{s}$\footnotemark[$*$] & $0.9739_{-0.0009}^{+0.0010}$ \\
                                  &  Flux$_{h}$\footnotemark[$\dagger$] & $7.948\pm0.011$ \\
                                  &  $\chi ^2 /$ d.o.f. & 540.28 / 86 \\
      \hline
      Band function\footnotemark[$\ddagger$]      & $\alpha$ & $-2.075_{-0.004}^{+0.002}$ \\
                                  &  $\beta$  & $> -2.387$ \\
                                  &  $E_{peak}$~(keV)  & $1378_{-287}^{+542}$ \\
                                  &  $\chi ^2 /$ d.o.f. & 429.25 / 85 \\
      \hline
      Broken power law      & $\Gamma_1$ & $-2.115\pm0.002$ \\
                                  &  $E_{break}$~(keV)  & $134_{-11}^{+7}$ \\
                                  &  $\Gamma_2$  & $-2.26\pm0.02$ \\
                                  &  Flux$_{s}$\footnotemark[$*$] & $0.9700_{-0.0008}^{+0.0009}$ \\
                                  &  Flux$_{h}$\footnotemark[$\dagger$] & $7.916\pm0.009$ \\
                                  &  $\chi ^2 /$ d.o.f. & 311.14 / 85 \\
      \hline
      \hline
\multicolumn{3}{@{}l@{}}{\hbox to 0pt{\parbox{100mm}{\footnotesize
 \par\noindent
 All errors are presented at 90\% confidence levels.\\
 \footnotemark[$*$] in units of $\times 10^{-8}$~erg cm$^{-2}$ s$^{-1}$ at 25--55~keV\\
 \footnotemark[$\dagger$]  in units of $\times 10^{-9}$~erg cm$^{-2}$ s$^{-1}$ at 50--100~keV\\
 \footnotemark[$\ddagger$]  Flux of the Band function cannot be derived as $\beta$ and $E_{peak}$ have too large errors.\\
 }\hss}}
   \end{tabular}
  \end{center}
\end{table}

\begin{table}
  \caption{Best-fit broken power-law model of individual spectra}\label{table3}
  \begin{center}
    \begin{tabular}{lllllllll}
      \hline
      ObsID & Date & MJD & $\Gamma_1$ & $E_{break}$~(keV) & $\Gamma_2$ & Flux$_{s}$\footnotemark[$*$] & Flux$_{h}$\footnotemark[$\dagger$] & $\chi^2$ \footnotemark[$\ddagger$]\\ \hline
      100023010 & 2005/09/15 	& 52628 & $-2.093^{+0.014+0.007}_{-0.010-0.004}$ & $120_{-25-10}^{+22+5}$ & $-2.30^{+0.08+0.04}_{-0.10-0.10}$ & $1.005_{-0.006}^{+0.005}$ & $8.32_{-0.06}^{+0.06}$ & 63.86 \\
      100023020 & 2005/09/15 	& 53628 & $-2.093^{+0.020+0.004}_{-0.015-0.005}$ & $92_{-20-6}^{+29+8}$ & $-2.21^{+0.03+0.03}_{-0.06-0.08}$ & $1.000_{-0.006}^{+0.006}$ & $8.28_{-0.1}^{+0.05}$ & 78.39 \\
      101004010 & 2006/03/30 	& 53824 & $-2.125^{+0.018+2.107}_{-0.010-0.005}$ & $132_{-56-76}^{+38+29}$ & $-2.26^{+0.09+0.03}_{-0.14-0.14}$ & $1.000^{-0.006}_{+0.006}$ & $8.11_{-0.07}^{+0.06}$ & 88.45 \\
      101004020 & 2006/04/04  	& 53829 & $-2.109^{+0.008+0.000}_{-0.008-0.005}$ & $141_{-19-0}^{+16+1}$ & $-2.35^{+0.09+0.00}_{-0.11-0.12}$ & $0.993_{-0.005}^{+0.005}$ & $8.13_{-0.05}^{+0.05}$ & 109.82 \\
      101003010 & 2006/04/05  	& 52830 & $-2.094^{+0.005+0.011}_{-0.005-0.004}$ & $135_{-15-34}^{+12+1}$ & $-2.28^{+0.05+0.09}_{-0.06-0.09}$ & $0.981_{-0.003}^{+0.003}$ & $8.11_{-0.03}^{+0.03}$ & 122.97 \\
      101010010 & 2006/09/05  	& 53983 & $-2.106^{+0.012+0.000}_{-0.010-0.004}$ & $92_{-17-0}^{+17+304}$ & $-2.18^{+0.03+4.96}_{-0.03-0.05}$ & $0.982_{-0.005}^{+0.004}$ & $8.06_{-0.10}^{+0.04}$ & 96.36 \\
      102019010 & 2007/03/20 	& 54179 &$-2.093^{+0.007+0.003}_{-0.007-0.007}$ & $99_{-11-7}^{+20+12}$ & $-2.19^{+0.02+0.04}_{-0.03-0.10}$ & $0.949_{-0.003}^{+0.003}$ & $7.85_{-0.05}^{+0.03}$ & 85.28 \\
      103007010 & 2008/08/27 	& 54705 & $-2.120^{+0.006+0.008}_{-0.006-0.009}$ & $166_{-15-8}^{+19+8}$ & $-2.60^{+0.13+0.15}_{-0.21-0.40}$ & $0.987_{-0.004}^{+0.003}$ & $8.03_{-0.03}^{+0.03}$ & 119.03 \\
      103008010 & 2008/09/01  	& 54710 & $-2.105^{+0.005+0.013}_{-0.005-0.007}$ & $148_{-51-8}^{+14+1}$ & $-2.32^{+0.12+0.06}_{-0.08-0.14}$ & $0.979_{-0.003}^{+0.003}$ & $8.05_{-0.03}^{+0.03}$ & 139.51 \\
      104001010 & 2009/04/02  	& 54922 & $-2.103^{+0.007+0.000}_{-0.007-0.006}$ & $111_{-11-0}^{+13+0}$ & $-2.25^{+0.03+0.00}_{-0.04-0.07}$ & $0.969_{-0.003}^{+0.003}$ & $7.97_{-0.04}^{+0.04}$ & 107.22 \\
      104001070 & 2010/02/23 	& 55250 & $-2.123^{+0.011+0.000}_{-0.008-0.064}$ & $173_{-64-76}^{+171+0}$ & $-2.26^{+0.13+0.06}_{-0.79-0.38}$ & $0.949_{-0.005}^{+0.005}$ & $7.71_{-0.05}^{+0.05}$ & 86.83 \\
      105002010 & 2010/04/05  	& 55291 & $-2.124^{+0.006+0.000}_{-0.006-0.159}$ & $145_{-26-89}^{+27+0}$ & $-2.26^{+0.06+0.12}_{-0.09-0.19}$ & $0.950_{-0.003}^{+0.003}$ & $7.71_{-0.04}^{+0.03}$ & 130.47 \\
105029010 & 2011/03/21  & 55641 & $-2.133^{+0.005+0.011}_{-0.005-0.005}$ & $247_{-19-143}^{+30+0}$ & $-1.44^{+0.35+0.20}_{-0.22-0.57}$ & $0.958_{-0.003}^{+0.003}$ & $7.73_{-0.03}^{+0.03}$ & 111.70 \\
106012010 & 2011/09/01 & 55805 & $-2.117^{+0.007+0.005}_{-0.007-0.006}$ & $108_{-15-93}^{+34+475}$ & $-2.20^{+0.03+0.06}_{-0.05-0.07}$ & $0.965_{-0.003}^{+0.003}$ & $7.87_{-0.05}^{+0.04}$ & 97.46 \\
106013010 & 2012/02/28  & 55986 & $-2.173^{+0.023+0.034}_{-0.035-0.000}$ & $45_{-12-0}^{+10+242}$ & $-2.11^{+0.01+1.15}_{-0.01-0.28}$ & $0.954_{-0.005}^{+0.006}$ & $7.69_{-0.06}^{+0.09}$ & 100.32 \\
106014010 & 2012/03/14  & 56001 & $-2.125^{+0.005+0.008}_{-0.005-0.009}$ & $166_{-15-6}^{+17+6}$ & $-2.54^{+0.11+0.13}_{-0.15-0.28}$ & $0.970_{-0.003}^{+0.003}$ & $7.87_{-0.03}^{+0.03}$ & 103.74 \\
106015010 & 2012/03/26  & 56012 & $-2.133^{+0.007+0.009}_{-0.007-0.010}$ & $173_{-17-8}^{+18+4}$ & $-2.74^{+0.18+0.16}_{-0.25-0.34}$ & $0.970_{-0.004}^{+0.004}$ & $7.83_{-0.04}^{+0.04}$ & 128.49 \\
      \hline
\multicolumn{9}{@{}l@{}}{\hbox to 0pt{\parbox{180mm}{\footnotesize
 \par\noindent
 Values of the first errors of $\Gamma_1$, $E_{break}$ and $\Gamma_2$ are statistical errors at 90\% confidence levels.
 Each sum of two values of errors represents
 the sum of statistical and systematic errors of the GSO NXB (1\%; see text) at 90\% confidence levels.\\
 \footnotemark[$*$] in unit of $\times 10^{-8}$~erg cm$^{-2}$ s$^{-1}$ at 25--55~keV\\
 \footnotemark[$\dagger$]  in unit of $\times 10^{-9}$~erg cm$^{-2}$ s$^{-1}$ at 50--100~keV\\
 \footnotemark[$\ddagger$]  d.o.f. = 85.
 }\hss}}
   \end{tabular}
  \end{center}
\end{table}

\begin{figure}
  \begin{center}
    \FigureFile(160mm,120mm){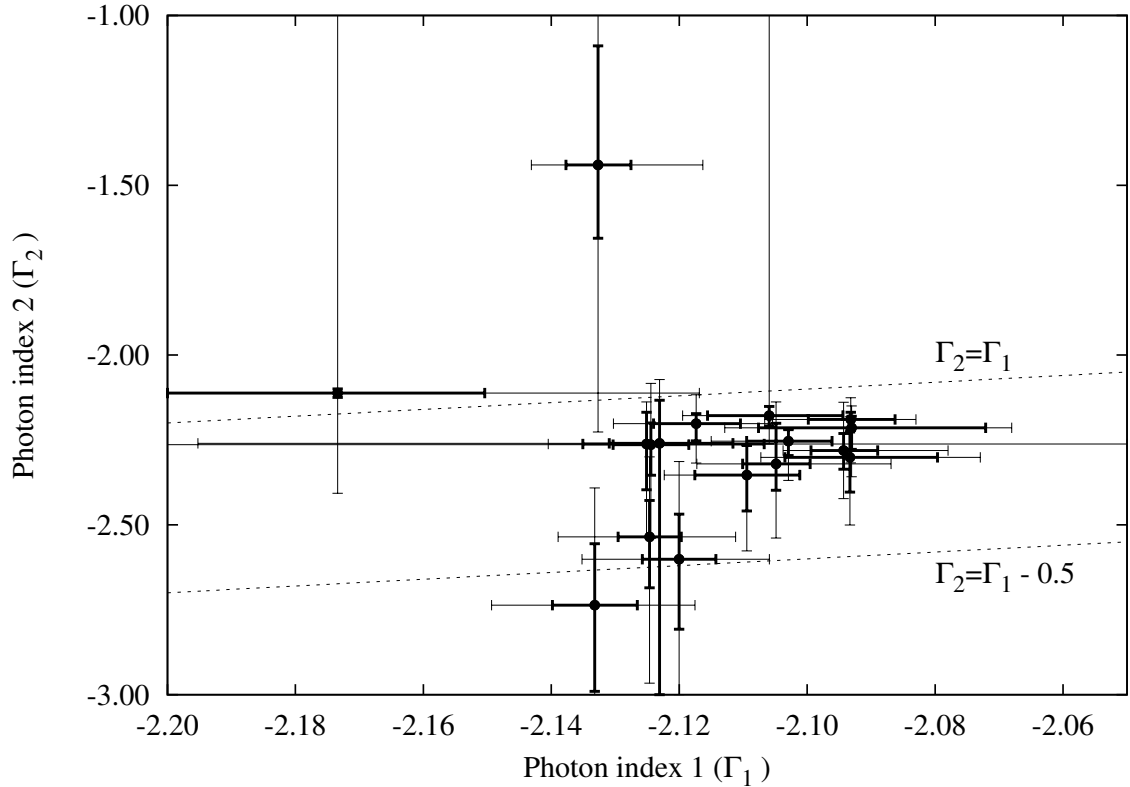}
  \end{center}
  \caption{
$\Gamma_1$ vs. $\Gamma_2$ plots of the best-fit broken power-law models for individual spectra.
The small error bars (bold lines) represent statistical errors
and the large error bars (fine lines) represent systematic errors of the NXB uncertainty (see text).
All errors are on a 90\% confidence level.
}\label{figure6}
\end{figure}

\begin{figure}
  \begin{center}
    \FigureFile(120mm,90mm){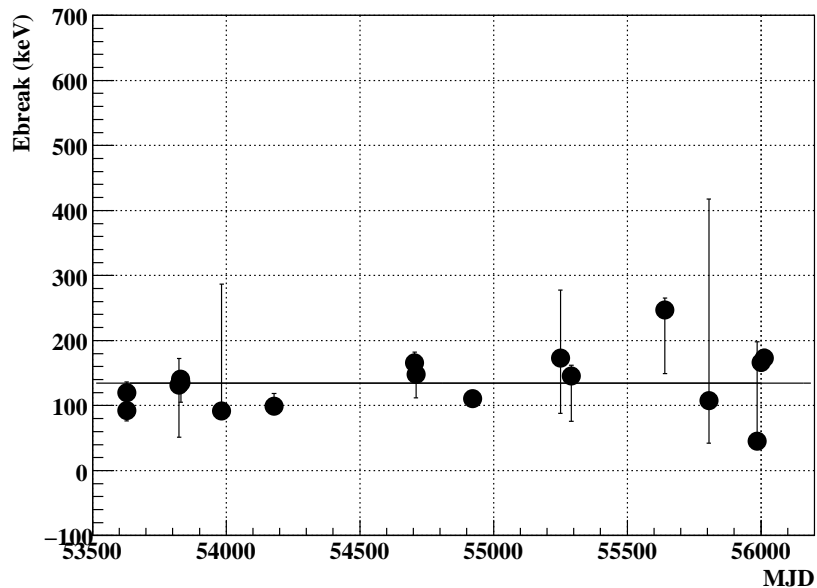}
  \end{center}
  \caption{
Time (MJD) vs. $E_{break}$ of the best-fit broken power-law models for individual spectra.
The error bars represent statistical errors plus systematic errors of the NXB uncertainty (see text).
All errors are on a 1$\sigma$ level.
The solid line represents the best-fit constant function.
}\label{figure7}
\end{figure}

\begin{figure}
  \begin{center}
    \FigureFile(120mm,90mm){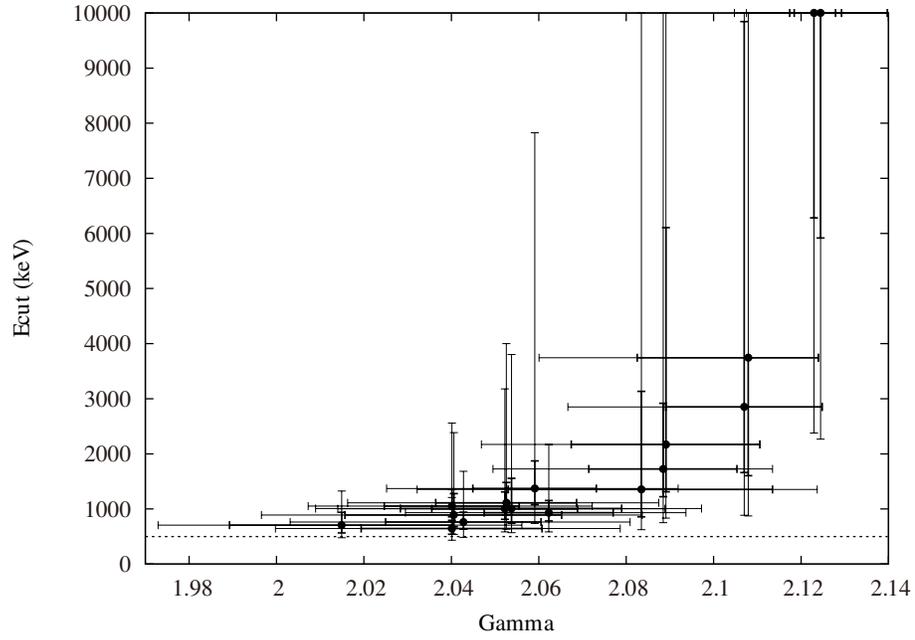}
  \end{center}
  \caption{
$\Gamma$ vs. $E_{cut}$ of the best-fit cutoff power-law models for individual spectra.
The dashed line shows $E_{cut} =$ 500~keV.
The small error bars (bold lines) represent statistical errors;
the large error bars (fine lines) represent systematic errors of the NXB uncertainty (see text).
All errors are on a 90\% confidence level.
}\label{figure8}
\end{figure}


\begin{figure}
  \begin{center}
    \FigureFile(80mm,80mm){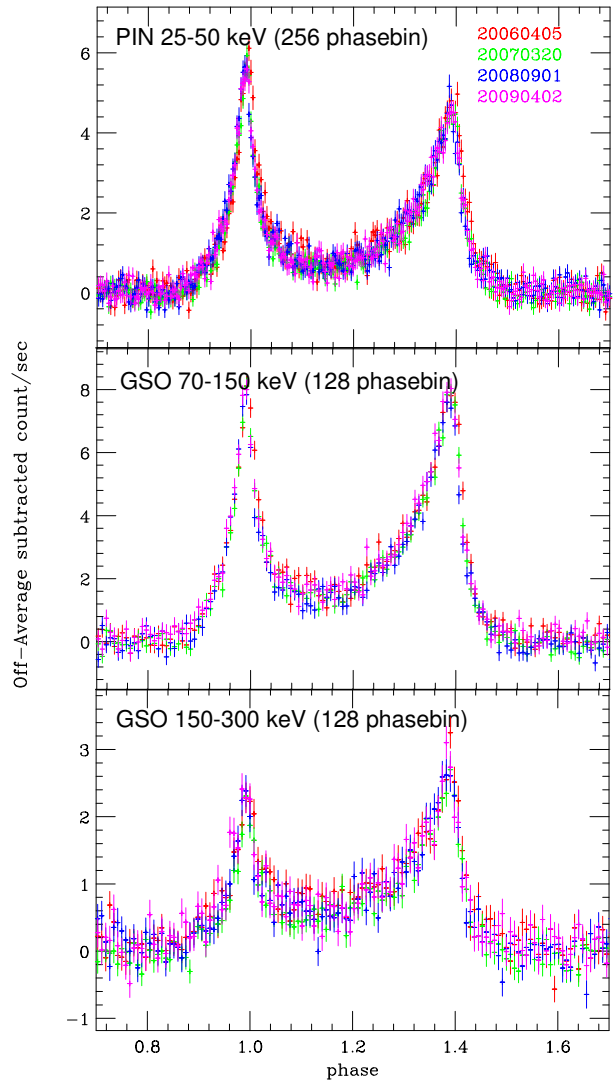}
  \end{center}
  \caption{Pulse profiles of the Crab pulsar
  observed with the HXD nominal position.
  Vertical axes show the count rates with subtraction of  the average Off phase values.
  The error bars represent statistical errors (1$\sigma$).
  }
	\label{figure9}
\end{figure}

\begin{figure}
  \begin{center}
    \FigureFile(100mm,80mm){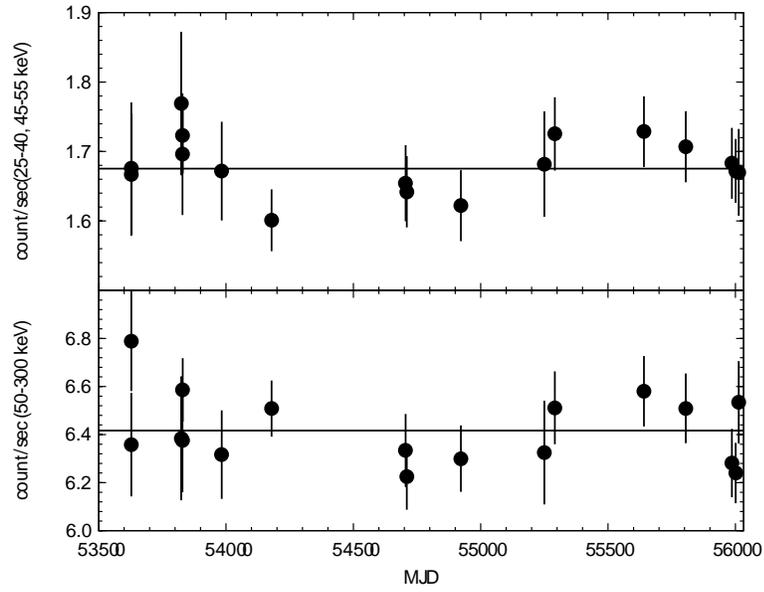}
  \end{center}
  \caption{Count rates of the Pulsed components of the Crab pulsar in 25--40, 45--55 keV of the PIN (top) and in 50--300 keV of the GSO (bottom).
The dashed line in each panel represents the best-fit constant function in each energy band.
All errors are on a 90\% confidence level.
}\label{figure10}
\end{figure}

\end{document}